\documentstyle[prl,aps,epsfig]{revtex}
\tighten
\draft
\begin{document}
\twocolumn[\hsize\textwidth\columnwidth\hsize\csname 
@twocolumnfalse\endcsname

\title{A vortex description of the first-order phase transition in type-I superconductors}
\author{Lu\'{\i}s M. A. Bettencourt$^1$ and Greg J. Stephens$^2$}
\address{$^1$Center for Theoretical Physics, Massachusetts Institute of Technology,
Bldg. 6-308, Cambridge MA 02139}
\address{$^2$Theoretical Division T-6 MS B288, Los Alamos National Laboratory,
Los Alamos NM 87545}

\date{\today}
\maketitle

\begin{abstract}
Using both analytical arguments and detailed numerical evidence we show 
that the first order transition in the type-I 2D Abelian Higgs model can be 
understood in terms of the statistical mechanics of vortices, 
which behave in this regime as an ensemble of attractive particles. 
The well-known instabilities of such ensembles are shown to be 
connected to the process of phase nucleation. By characterizing the equation of state 
for the vortex ensemble we show that the temperature for the onset of a clustering 
instability  is in qualitative agreement with the critical temperature. 
Below this point the vortex ensemble collapses to a single cluster, which is 
a non-extensive phase, and disappears in the absence of net topological charge. 
The vortex description provides a detailed mechanism for the 
first order transition, which applies at arbitrarily weak type-I and is gauge invariant, 
unlike the usual field-theoretic considerations, which rely on asymptotically large 
gauge coupling.  
\end{abstract}

\pacs{PACS Numbers : 74.55.+h, 11.27.+d, 74.40.+k, 05.70.Fh  \hfill MIT-CTP-3169, LAUR-01-4286}

\vskip2pc]
    
The role of topological excitations in the dynamics and thermodynamics of  
gauge field theories is a subject of great promise ranging from the understanding 
of vortex phases in superconductors, necessary for practical applications, 
to the clarification of the mechanisms of charge confinement in non-Abelian 
gauge theories, such as quantum chromodynamics. 

Topological excitations are important as finite energy vehicles of disorder. Thus, phase transitions 
between a state of long range (e.g. magnetic) order and disorder can sometimes be understood as 
the proliferation of topological excitations, each bringing about disorder 
comparable to its size \cite{KBook}. 
This is true in the XY model in two spatial dimensions (2D), 
which displays a Kosterlitz-Thouless (KT) transition to a disordered state due to vortex pair 
unbinding \cite{KT}.  The second order transition in the 2D Ising model can also be formulated in terms 
of domain wall percolation. Furthermore there is evidence that the second order transition in the 
3D XY universality class is associated with vortex string proliferation \cite{KBook,3DXY}.

In this letter we show how a {\em first-order} (discontinuous) phase transition in a simple 
gauge theory can be understood in terms of topological excitations and provide detailed numerical 
evidence in support of this view. Our results suggest a dual vortex gas picture of the transition,
in analogy to the KT case. In the type-I Abelian gauge theory, however, vortices have 
attractive interactions, leading to characteristic metastability and collapse.

For its simplicity and close relationship to the XY model we study the 
Abelian-Higgs (or Landau-Ginzburg) model in 2D. 
The Lagrangian density $\cal L$ is
\begin{eqnarray}
{\mathcal{L}}= -{1 \over 4}F_{\mu\nu}F^{\mu\nu} +{1 \over 2} |D_\mu \phi|^2 
-{\lambda \over 8} \left( |\phi|^2 - v^2 \right)^2,
\label{AHMLagr}
\end{eqnarray}
where $\phi$ is a complex scalar field, $F_{\mu\nu}=\partial_\mu A_\nu-\partial_\nu A_\mu$ 
is the field strength for the gauge potential $A_\mu$ and 
$D_\mu =\partial_\mu + i e A_\mu$. We focus on the regime where $e^2$ 
is larger than the scalar coupling $\lambda$. This model describes the 
long-wavelength behavior of an ideal type-I superconductor. 

The standard argument \cite{HLM} for a first order transition in gauge+scalar 
field theories relies on large $e$, for which the gauge field is very massive and can be integrated out. 
This is only justifiable as $\kappa \equiv \sqrt{\lambda}/e \rightarrow 0$, 
as it requires a separation of scales between `heavy' gauge degrees of freedom, 
which do not participate in the transition dynamics,  and `light' 
scalar field fluctuations.  The result is a `free energy' $F[\phi]$ that is both non-convex and, 
generally, gauge dependent. Nevertheless $F[\phi]$ yields the correct qualitative 
picture for certain aspects of the transition.

A description of the critical system in terms of gauge invariant 
degrees of freedom, for arbitrary $\kappa < 1$, is therefore desirable and may shed light on 
the mechanism of the transition. While vortices exist only as fluctuations at high temperature
they become the only long lived magnetic excitations of (\ref{AHMLagr}) at low temperatures. 
Moreover, arbitrarily low energy excitations can be produced by the superposition of vortices 
and anti-vortices. These arguments suggest that vortices are relevant degrees of freedom at criticality. 
In type-I, vortices attract each other independently of the sign of their quantized flux 
(topological charge) \cite{Bogomolny}. Thus Abrikosov vortex lattices are not formed in applied magnetic 
fields. Instead type-I superconductors form a non-extensive multi-winding vortex, restoring the normal
phase at its core.  Between the normal and superconducting phases type-I superconductors exhibit 
a first-order transition.

Vortices are  radial, static classical solutions obeying
\begin{eqnarray}
&& \nabla^2_r \sigma - \left[ \left( e A_\theta - {n \over r} \right)^2 + {\lambda \over 2} 
(\sigma^2 -v^2 ) \right] \sigma = 0, \\
&& \nabla \time \nabla \times A_\theta + e^2 \sigma^2 \left(  A_\theta - {n \over e r} \right)  = 0,
\end{eqnarray}  
in temporal gauge $A_0=0$, where $\phi = \sigma(r) e^{i n \theta}$. 
$\theta$ is the polar angle, $n$ is an integer, 
$ A_\theta(r) = n/er - a(r)/r$, with boundary conditions 
$\sigma (r=0) = a(r=0)= 0$,  $\sigma (r \rightarrow \infty) = v, \ a(r \rightarrow \infty) 
\rightarrow n/e$.
For $r$ larger than the core size the vortex behaves like a point source
for massive scalar and magnetic fields. Then the vortex profiles can be 
written \cite{Bettencourt}
\begin{eqnarray}
&& \sigma(r) = v - f (r); \qquad f(r)=  a_S~v~q~K_0(m_S r), \label{pointchargef} \\
&& A_\theta (r)  - {n \over e r} = - a_G~v~m~K_1(m_G r), 
\label{pointchargeA}
\end{eqnarray}  
where $m_S=\sqrt{\lambda} v$, $m_G= e v$,  $a_S,~a_G$ are dimensionless constants, and $K_i$, modified 
Bessel functions of order $i$. The profiles (\ref{pointchargef}-\ref{pointchargeA}) correspond 
to  Yukawa (massive) charges in 2D, in contrast to the familiar Coulomb logarithmic 
vortex solutions of the the 2D XY model.

In the XY model the importance of topological charges to the phase transition 
is demonstrated by rewriting the partition function in terms of vortex degrees of 
freedom \cite{Coulomb}. Unfortunately because the Abelian Higgs system is non-Gaussian it is 
impossible to perform an exact dual transformation to a partition function written exclusively 
in terms of a one and two body vortex terms. It is nevertheless possible to perform this 
re-writing approximately.

We begin with a superposition ansatz for an {\em arbitrary} number $N$ 
of vortices, by constructing scalar and gauge vortex fields
centered at $N$ different loci $r_i$, $i \in \left\{1,N \right\}$
\begin{eqnarray}
&&\phi(r,r_1,...r_N) = {\phi(\vert r-r_1\vert) ... 
\phi(\vert r-r_N\vert) \over v^{N-1}}, \label{ansatzphi} \\
&&\vec A_i(r,r_1,...r_N) = \vec A_1(\vert r-r_1\vert) + \ldots 
+ \vec A_N(\vert r-r_N\vert). \label{ansatzA} 
\end{eqnarray}
This ansatz is exact when the vortices 
are all coincident or all widely separated.
Substituting (\ref{ansatzphi}-\ref{ansatzA}) into the static part of the Hamiltonian gives 
\begin{eqnarray}
H = \sum_i \epsilon_i + && \sum_{<i,j>} \left[ m_i(x) V_G(\vert x-y \vert) m_j(y) \right. \\  
&& \left. \qquad + q_i(x) V_S(\vert x-y \vert) q_j(y) \right] + \ldots , \nonumber 
\end{eqnarray}
where the terms not shown correspond to 3 and 4-body effects, which 
are negligible in a low-density vortex ensemble. 
The charges $m_i (x)= \pm n \delta(x)$  are integers of either sign, corresponding to 
quanta of magnetic flux, whereas $q_i= \vert m_i \vert$ is always positive \cite{Bettencourt,Speight}. 
For well-separated vortices the potentials are 
$V_G(r) = a_G~ v_{\rm eq}^2~ K_0 ( m_G r ) $, $V_S(r) = - a_S~ v_{\rm eq}^2~ K_0 ( m_S r )$ 
\cite{Bettencourt}. $v_{\rm eq}$ is a measure of $\sigma(T)$ and $a_S,~a_G$ are weakly varying with 
the couplings and have been computed numerically by Speight \cite{Speight}: for $\kappa < 1$, 
$a_S > a_G$. Then the 2-body potential for a pair of like-charge vortices is
\begin{eqnarray}
V (r) \simeq  - q_i q_j v_{\rm eq}^2 \left[  a_S^2 K_0 ( m_S r ) -   a_G^2 K_0 ( m_G r ) \right],  
\label{vortexpotential}
\end{eqnarray}
\noindent demonstrating that in type-I, when $\kappa < 1$, the scalar (attractive) part of 
the potential dominates the interaction.
\begin{figure}
\begin{minipage}[t][9.6cm][t]{9cm} 
\epsfig{figure=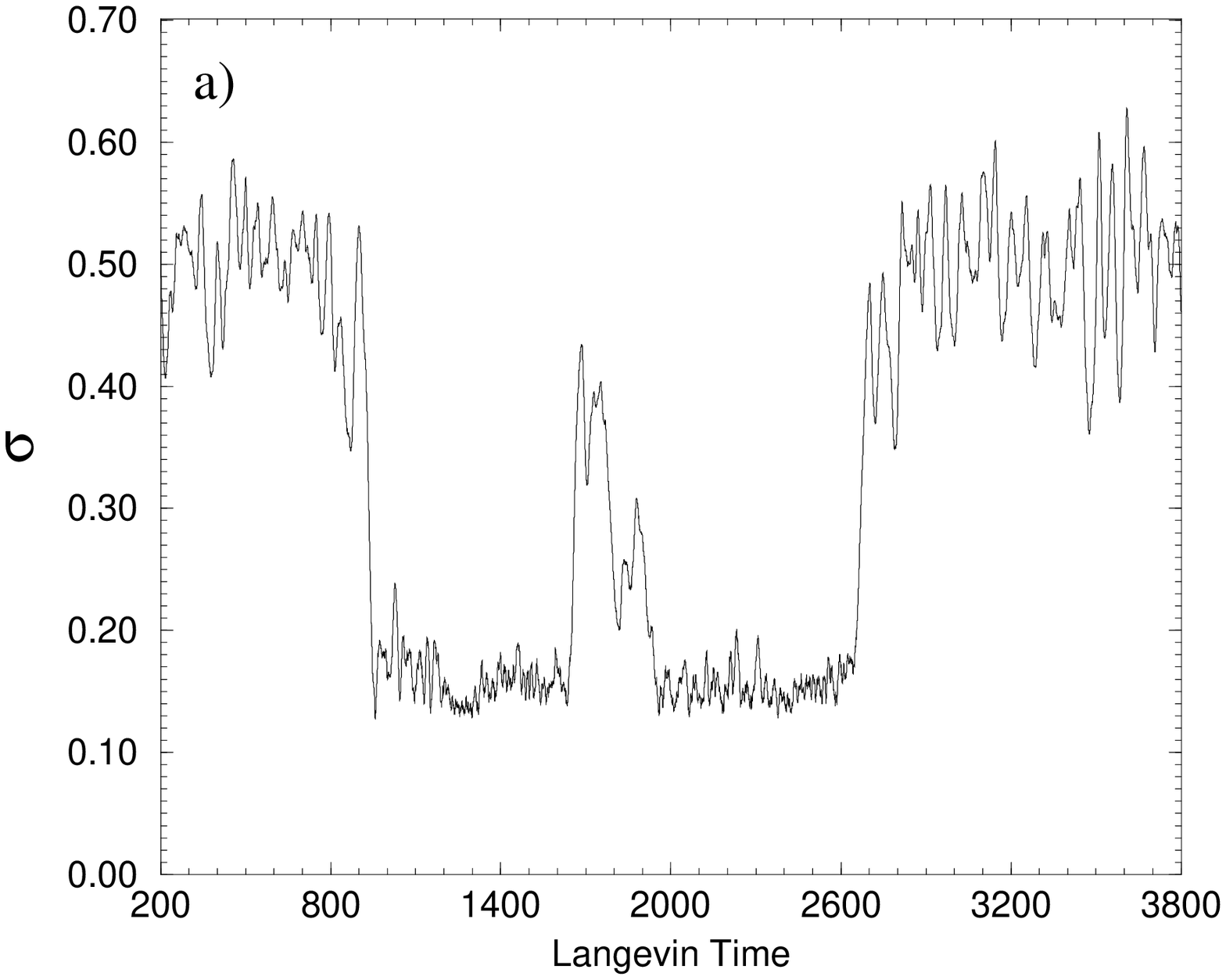,width=3.2in,height=2.0in}
\begin{minipage}[t][4.3cm][b]{9cm} 
\epsfig{figure= 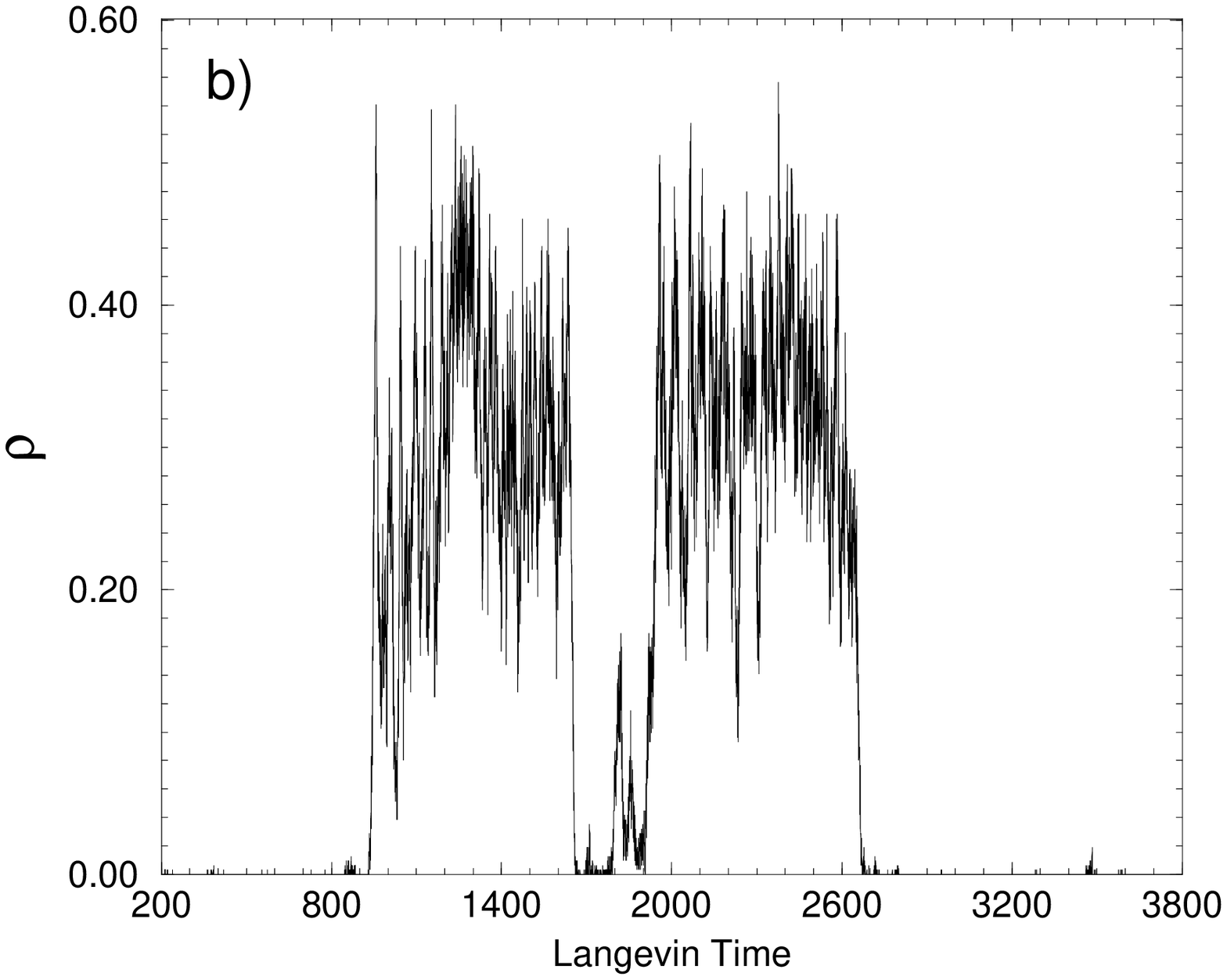,width=3.2in,height=2.0in} 
\end{minipage}
\end{minipage} 
\caption{a) A characteristic time evolution of at $T=0.01844$, close to phase 
coexistence. $\sigma$ jumps abruptly between the normal and superconducting phases. 
b) The total vortex density $\rho$ for the same trajectory as in a). 
$\rho$ is a disorder parameter vanishing in the superconducting phase.} 
\label{fig1}
\end{figure}

The statistical mechanics of particles with a finite-range, soft-core 
attractive potential was studied in several simple settings \cite{attractivegas}, 
aimed at elucidating thermodynamic gravitational instabilities. Unfortunately 
the Yukawa gas of (\ref{vortexpotential}), was not among these.
Nevertheless this class of systems share important qualitative properties; namely
they always display a first-order transition between an (almost) ideal gas state at 
high $T$ and a non-extensive clustered phase at low $T$ \cite{attractivegas}. 
The latter is {\it not} a thermodynamic phase.  It consists of a single 
bound cluster containing most of the particles.
Qualitatively this clustering transition occurs at $T_{\rm cl}$, 
such that the kinetic energy equals the interaction energy per particle.  
This gives a rough estimate of the vortex gas clustering temperature \cite{attractivegas}, 
$T_{\rm cl} \sim \epsilon/2 = 0.019$ where $\epsilon$ measures the strength 
of the interaction.  We estimate $\epsilon$ from \cite{Jacobs}, by averaging the strength of
vortex-vortex and vortex-antivortex pair interactions,
$\epsilon= (\epsilon_{vv} + \epsilon_{v \bar v})/2$, where $\epsilon_{vv}\simeq 0.62 v_{\rm eq}^2$ and 
$\epsilon_{v \bar v} = 3.80 v_{\rm eq}^2$.  Vortex interactions are softened by a small order parameter,
$v_{\rm eq}=\sigma(T_c^+) \simeq 0.13$. 

The equation of state for the almost ideal (low density) gas of vortices at high $T$  can 
be computed by standard cluster expansion methods. For simplicity we model the attractive potential 
as a square well with a strength $\epsilon$ and an interaction length $l$. Then 
\begin{eqnarray}
&& P \simeq \rho T \left(1 - \rho B_2\right),
\quad B_2=\frac{\pi l^2}{2} \left (\exp{\frac{\epsilon}{T}}-1 \right),
\label{eqstate}
\end{eqnarray}
where we neglected terms proportional to $\rho^n, n\geq 3$.
The correction to the ideal gas behavior is negative as expected for an attractive potential.
For high $T$, $B_2$ vanishes. Interactions are most important
at low $T$ and the pressure $P$ vanishes at $T_{\rm cl}$:
\begin{eqnarray}
T_{\rm cl} \simeq \frac {\epsilon}  {\ln{\left(1+\frac{2} {\rho \pi l^2} \right)}}.
\end{eqnarray} 
Using $\rho \pi l^2 \simeq 0.37 = \sigma(T_c^-) - \sigma(T_c^+)$, 
and $\epsilon$ as above we obtain $T_{\rm cl} \simeq 0.020$. 
Both estimates of $T_{\rm cl}$ are compatible with the measured $T_c$ and
coincide in the limit $\epsilon/T \ll 1$ and $\rho \pi l^2 \simeq 1$.   
\begin{figure}
\begin{minipage}[t][9.3cm][t]{9cm} 
\epsfig{figure=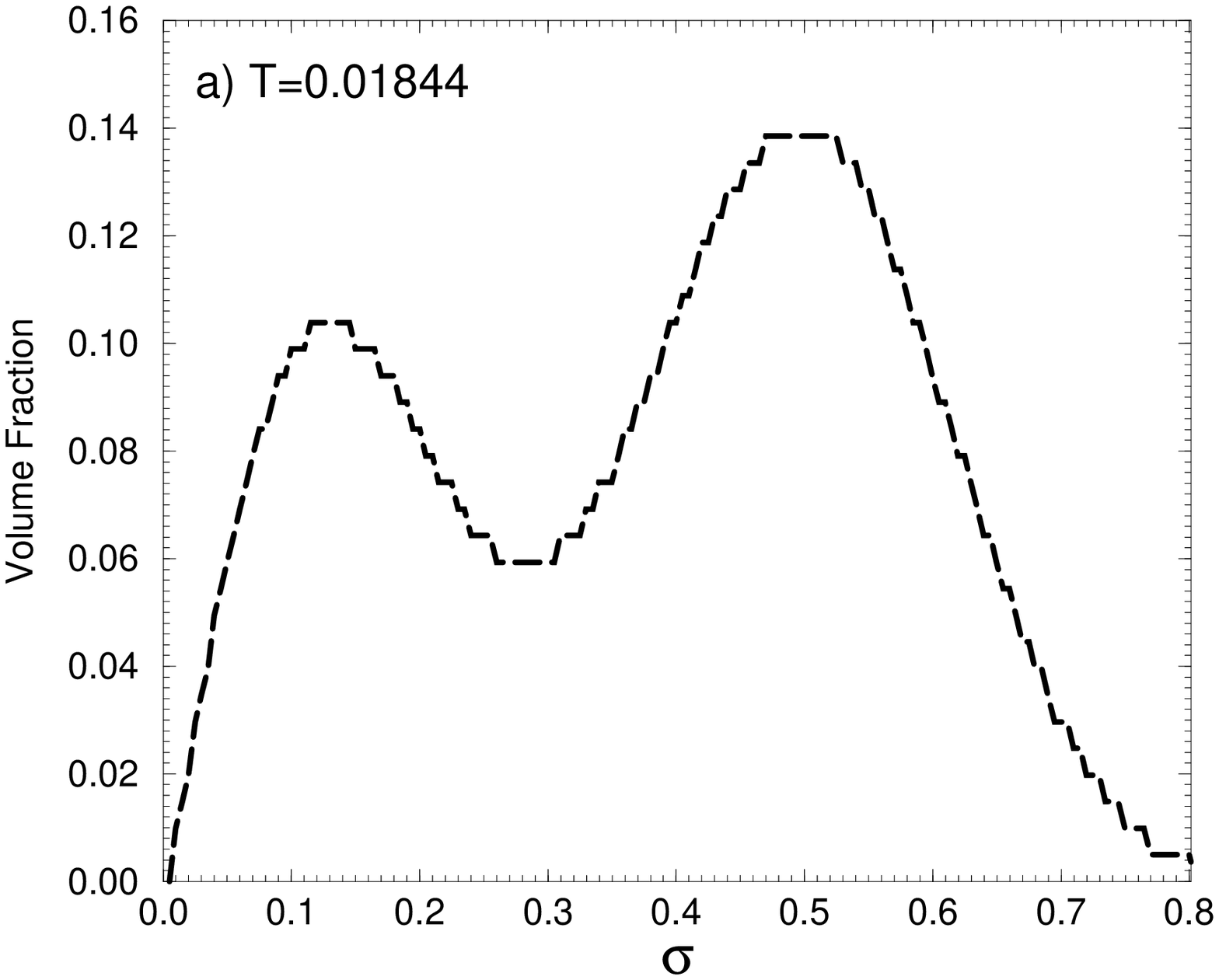,width=3.2in,height=2.0in}
\begin{minipage}[t][4.2cm][b]{9cm} 
\epsfig{figure=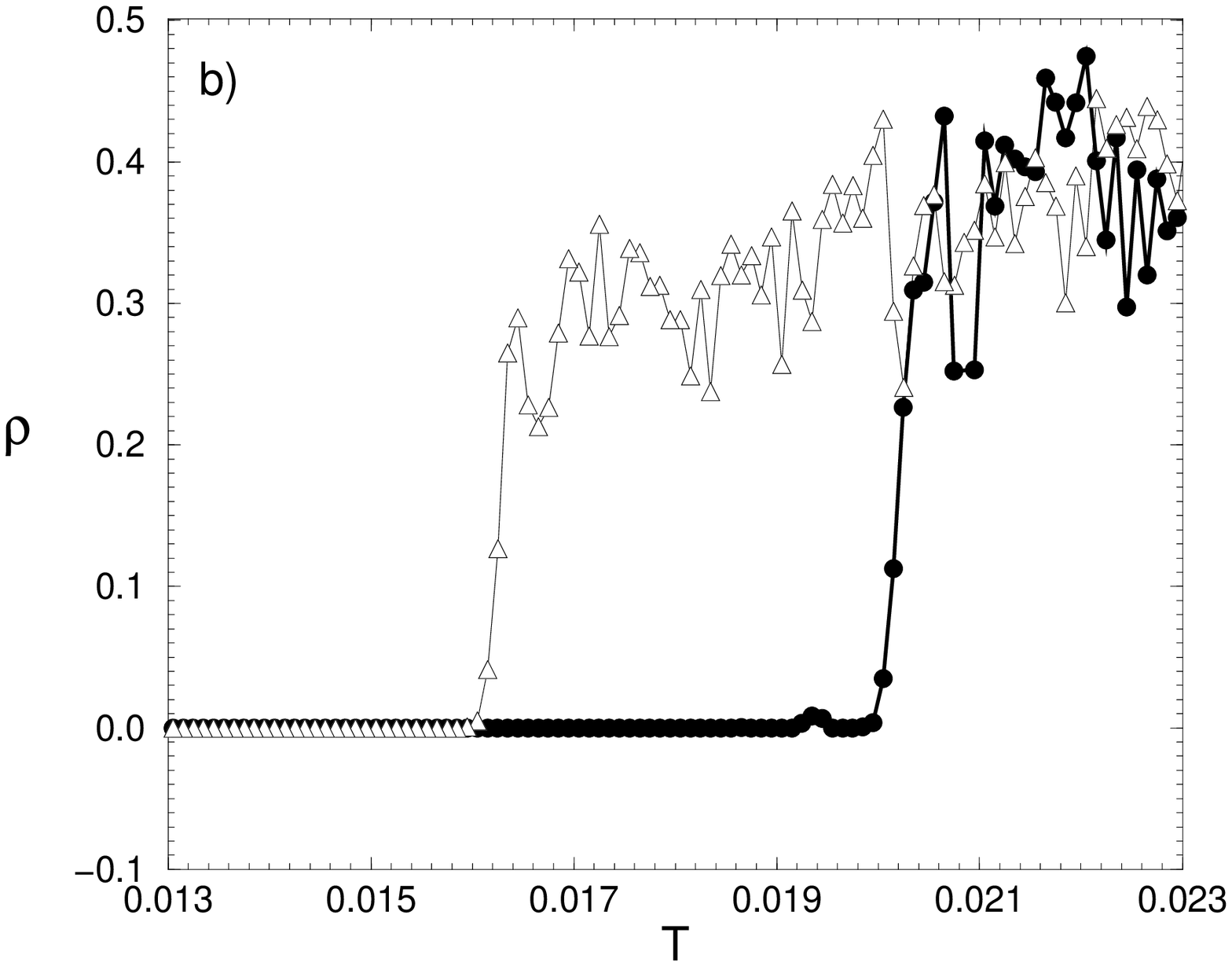,width=3.2in,height=2.0in} 
\end{minipage} 
\end{minipage}
\caption{ a) The probability distribution of $\sigma$ at $T=0.01844$, in the critical 
region, showing coexistence of the normal and superconducting phases. 
b) A hysteresis loop for the vortex density $\rho$ obtained by  
cooling (triangles) and heating (circles) the system through the critical region.}
\label{fig2}
\end{figure}

These estimates of $T_{\rm cl}$ are very qualitative in nature. 
For our choice of parameters the exact 2-vortex potential is not known. Moreover the virial expansion is
notoriously unreliable near a transition, especially in the presence of clustering. 
We plan to examine the quantitative behavior of the vortex 
particle ensemble, through a direct thermodynamic study using either the exact intervortex 
potential or the Yukawa gas interaction of (\ref{vortexpotential}).
  
Nevertheless, a vortex ensemble picture leads to several qualitative predictions
i) $\rho$ will behave as a disorder parameter, vanishing discontinuously 
in the superconducting phase,  ii) the vortex ensemble must show signs of a clustering instability 
in the metastable phase and iii) multi-charged vortices must be visible under sudden 
non-equilibrium cooling. 

To test these predictions we consider the field evolution in contact 
with a heat bath, given by a system of Langevin field equations.
Gauge invariance demands that the evolution preserves Gauss' law. 
This constraint still allows for several classes of dynamical equations \cite{Krasnitz}, 
characterized by different gauge invariant stochastic generators. We choose 
the simple set $\{|\phi|^2,\vec{E}\}$, leading to  
\begin{eqnarray}
\label{eq-langevin} 
&& \partial_t \pi_a = \left[ \nabla^2 - e^2 \vert A \vert^2   
- {\lambda \over 2} (\vert \phi \vert^2 -1) \right]\phi_a 
- 2 e \epsilon_{ab} A^i \partial_i \phi_b \nonumber \\
&& \qquad \qquad \qquad \qquad \qquad -2 \phi_i \left[ 
\eta_s \partial_t \vert \phi \vert^2 + \Gamma_s\right], \label{eqmotion} \\
&&\partial_t \phi_i = \pi_i, \nonumber \\
&& \partial_t E_i = (\nabla \times B)_i +J_i, 
\ \ \ J_i\equiv- e^2 \vert \phi \vert^2 A_i  
- e \epsilon_{ab} \phi_a \partial_i \phi_b,    \nonumber \\
&&\partial_t A_i = E_i + \eta_g \partial_t E_i  + \Gamma_g, \nonumber
\end{eqnarray}
with $E_i=\partial_t A_i$, $B=\vec \nabla \times \vec A$.  
The details of this choice are irrelevant to the state of 
canonical thermal equilibrium reached at long times.
The indices $a,b$ refer to the 2 real components of $\phi$, 
whereas $i$ is a spatial vector index. $\epsilon_{ij}$ is the 
totally anti-symmetric rank 2 tensor. The stochastic sources $\Gamma$ 
obey fluctuation-dissipation relations 
\begin{eqnarray}
\langle \Gamma_{s,g}(x,t) 
 \Gamma_{s,g}(x',t')\rangle= 2\eta_{s,g} T \delta(x-x')\delta(t-t'), 
\end{eqnarray}
with $\langle \Gamma_{s,g} \rangle =0$.
We choose $e=1.5,\: \lambda=0.1$, and solve a lattice (non-compact) version
of Eq.~(\ref{eqmotion}),  with $\eta_s=\eta_g=0.05$,  $dt=0.02$ and $dx=0.5$.

\begin{figure}
\begin{center}
\begin{minipage}[t][3cm][t]{8cm}
\epsfig{figure=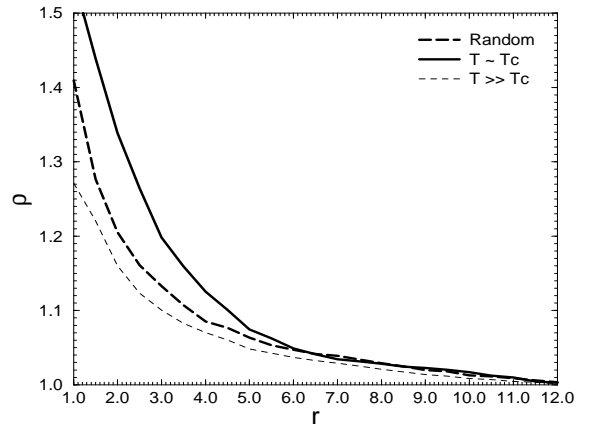,width=3.4in,height=2.6in}
\end{minipage}
\caption{The density of like-sign vortices $\rho(r)$ in a disc of radius $r$ around a vortex,
normalized by the average vortex density.  Plots correspond to the metastable phase 
(solid), the near-ideal gas phase at $T \gg T_c$ (short dashed) and to a random distribution 
(long dashed), with the same density as the metastable phase. 
The latter exhibits correlations typical of the subcritical percolation problem. 
Vortex clustering is maximal in the metastable phase, a premonitory sign of the dynamical 
clustering instability, and disappears for $T\gg T_c$.}
\label{fig3}
\end{center}
\end{figure}
Fig.~\ref{fig1} shows that a substantial $\rho \neq 0$ exists in the symmetric (normal) phase,
but that all vortices suddenly disappear when the system transits to the 
superconducting state, as shown by the spatial average of $\sigma=\vert \phi \vert$. 
Vortices in the normal phase are identified by their quantized  fluxes, a quantity that is manifestly 
gauge invariant.  The probability distribution of $\sigma$ close to $T_c$ is 
shown in Fig.~\ref{fig2}a. The double peak demonstrates coexistence of the 
two phases, characteristic of first order transitions. Fig.~\ref{fig2}b shows a hysteresis loop 
in $\rho$, obtained by slowly heating and then cooling through the critical region.

\begin{figure}
\begin{center}
\begin{minipage}[t][5.8cm][t]{5cm}
\epsfig{figure=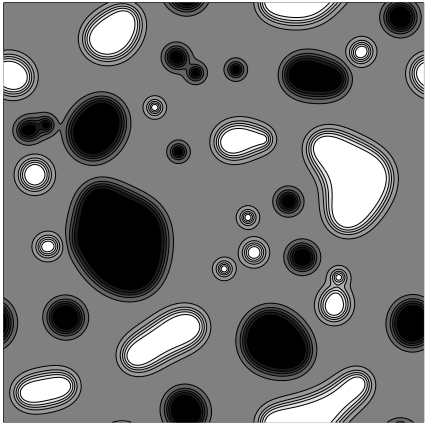,width=2.2in,height=2.2in}
\end{minipage}
\caption{Contours of magnetic flux after a fast temperature quench, 
showing the clustering of singly quantized vortices (smallest circular features) 
into large integer charge bound states. White (black), localized, regions denote 
(anti)vortices. The total collapse of the vortex ensemble was avoided 
due to fast cooling, which evaded the metastable region.}
\label{fig4}
\end{center}
\end{figure}

The metastability of the vortex ensemble is a consequence of the small 
probability for the formation of a large vortex cluster. While the free energy is 
lowered through attractive interactions in the volume, the spatial cluster boundary, 
where vortices are rarefied, is thermodynamically costly. 
Thus small clusters are subcritical 
and can exist in the metastable phase without leading to its collapse. 
Evidence for incipient clustering is shown in Fig.~\ref{fig3}, where we plot 
the radial density of like-charge vortices around another vortex. 
Clustering is maximal in the metastable region, and negligible at higher $T \gg T_c$.

While we argue that vortices are the relevant degrees of freedom at criticality, 
their profiles cannot be clearly observed in the normal phase, because vortices 
appear there only as transient fluctuations.  
Below $T_c$, where they could exist as well defined objects,
their ensemble collapses and vortices disappear in the absence of quantized net flux.
Fast quenches evading equilibrium in the metastable region do display well defined 
vortices and show striking evidence of their clustering, see  Fig.~\ref{fig4}.

In conclusion, we argued for a vortex description of the 
mechanism underlying the {\it first-order} transition in the 2D type-I Abelian Higgs model and 
provided detailed numerical evidence in support of this picture.  
Below $T_{\rm cl}$ the vortex ensemble becomes metastable and eventually collapses to 
a non-extensive thermodynamic phase and the system becomes a superconductor. 
The vortex interpretation of the transition is gauge invariant and 
does not require $e^2 \gg \lambda$,  unlike the field-theoretic arguments for a 
first-order transition. Instead, the attractive nature of the vortex potential is manifest even 
in the weakest type-I regime and  metastability and collapse are  inescapable. 

The vortex description must now be subjected to closer quantitative scrutiny by direct studies 
of particle ensembles. Still, we believe that its qualitative 
success bodes well for applications to 3D \cite{Kleinert}, where point vortices become 
lines, and may participate in interesting critical phenomena such as crystal melting and in 
cosmology.

We thank H.~Kleinert, N.~Rivier A.~Schakel and Z.~Tesanovic for useful comments on the manuscript.
This work was supported in part by the D.O.E. under research agreement $\#$DF-FC02-94ER40818.
Numerical work was done at the T/CNLS Avalon cluster at LANL.

\end{document}